\documentclass[12pt]{iopart}

\usepackage{iopams}  
\usepackage{graphicx} 

\begin{document}

\title[Dielectron continuum production at STAR  ]{Dielectron continuum production from $\sqrt{s_{NN}}$ = 200 GeV p + p and Au + Au collisions at STAR}

\author{Jie Zhao$^{1,2}$ (for the STAR collaboration)}

\address{1, Shanghai Institution of Applied Physics, CAS, Shanghai 201800, China 
}
\address{2, Lawrence Berkeley National Laboratory, Berkeley, CA 94720, USA 
}
\ead{zhaojie@sinap.ac.cn}
\begin{abstract}

We present the first STAR dielectron measurement in 200 GeV p + p and Au + Au collisions. Results are compared to hadron decay cocktails to search for vector meson in-medium modification in low mass region and  quark gluon plasma thermal radiation in the intermediate mass region. The $\omega \rightarrow e^{+}e^{-}$ spectra and the transverse mass distribution in the intermediate mass region are also discussed.  

\end{abstract}

\section{Introduction:}
Dilepton distributions have been proposed as one of the penetrating probes for hot and dense nuclear matter created in high-energy nuclear collisions. Due to their relatively small cross-sections with the hot/dense environment, dileptons bring us direct information of the created matter in such collisions. Since dileptons are created over all stages of heavy ion reactions, their sources vary as a function of kinematics. In the low mass region (LMR: $m_{ee}<$ 1.1 GeV/$c^{2}$), vector meson in-medium properties and directed photons can be studied through their di-leptonic decays, while in the intermediate mass region (IMR: 1.1 $< m_{ee} <$ 3 GeV/$c^{2}$) quark gluon plasma (QGP) radiation is expected to have a signification contribution at RHIC. In the high mass region (HMR: $m_{ee}>$  3 GeV/$c^{2}$), dileptons are mostly from heavy (charm and bottom) quark decays and Drell-Yan processes. As a result, the dilepton distributions, especially in the IMR and HMR, could provide new insights of early collision dynamics in heavy ion collisions.

With the completion of the full barrel time-of-flight detector\cite{TOF}, the electron identification has been significantly improved at STAR, especially in low momentum region. In this paper we will present the first STAR results on dielectron production in p + p and Au + Au collisions at $\sqrt{s_{NN}}$ = 200 GeV.

\section{Analysis}
The results are obtained from 200 GeV p + p and Au+Au collisions, taken in year 2009 and 2010, respectively. The main subsystems of the STAR detector\cite{TPC} used for this analysis are the Time Projection Chamber (TPC) and the Time-Of-Flight detector (TOF) with 2${\pi}$  azimuthal coverage at mid-rapidity.

In addition to track detection and momentum determination capabilities, the TPC provides particle identification for charged particles by measuring their ionization energy loss ($dE/dx$) in the TPC gas\cite{TPC}. The newly installed full barrel TOF detector provides the particle velocity ($\beta$) information\cite{TOF}. In this analysis we combine TPC $dE/dx$ and TOF $\beta$ information to identify electrons. The electron purity in p + p analysis is  $\sim 99\%$, and $\sim 97\%$ in minimum-bias Au + Au analysis.

\begin{figure}[htbp!]
    \centering 
    \includegraphics[width= 6.8cm]{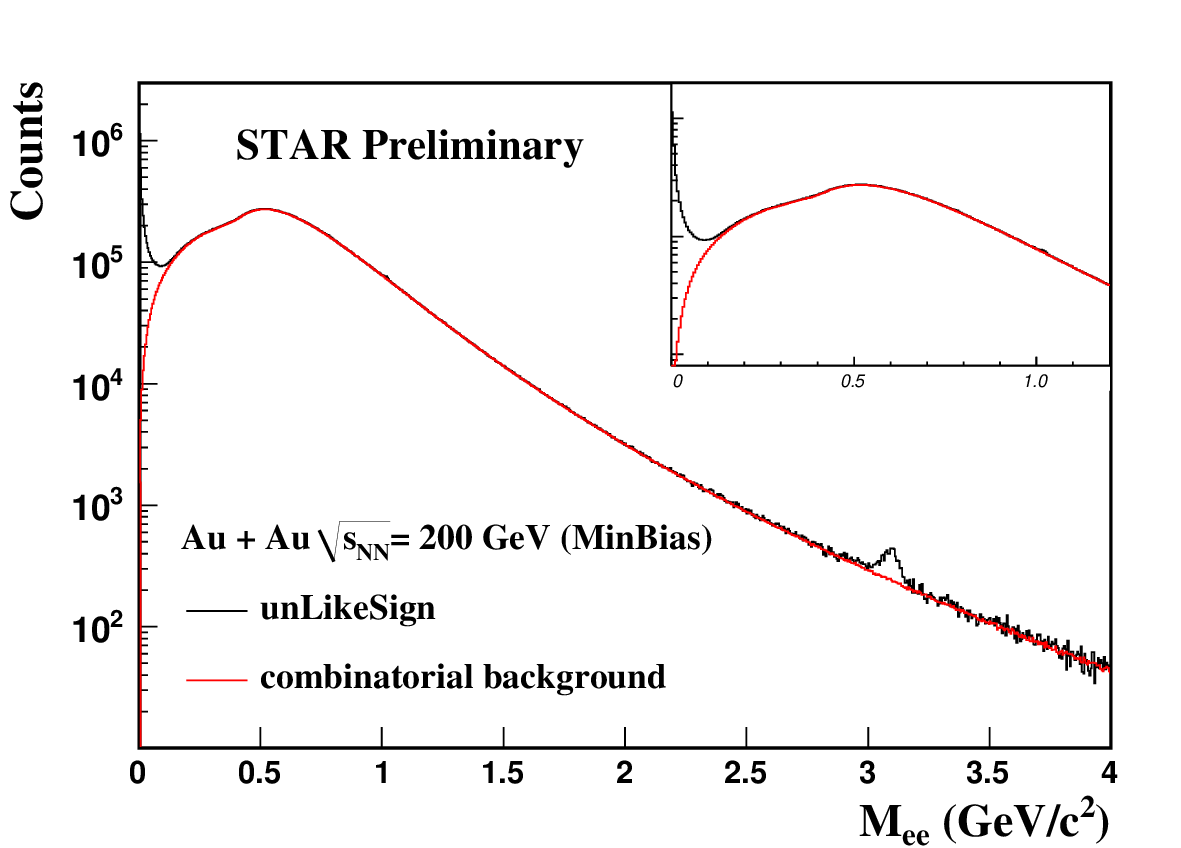} 
     \includegraphics[width= 6.8cm]{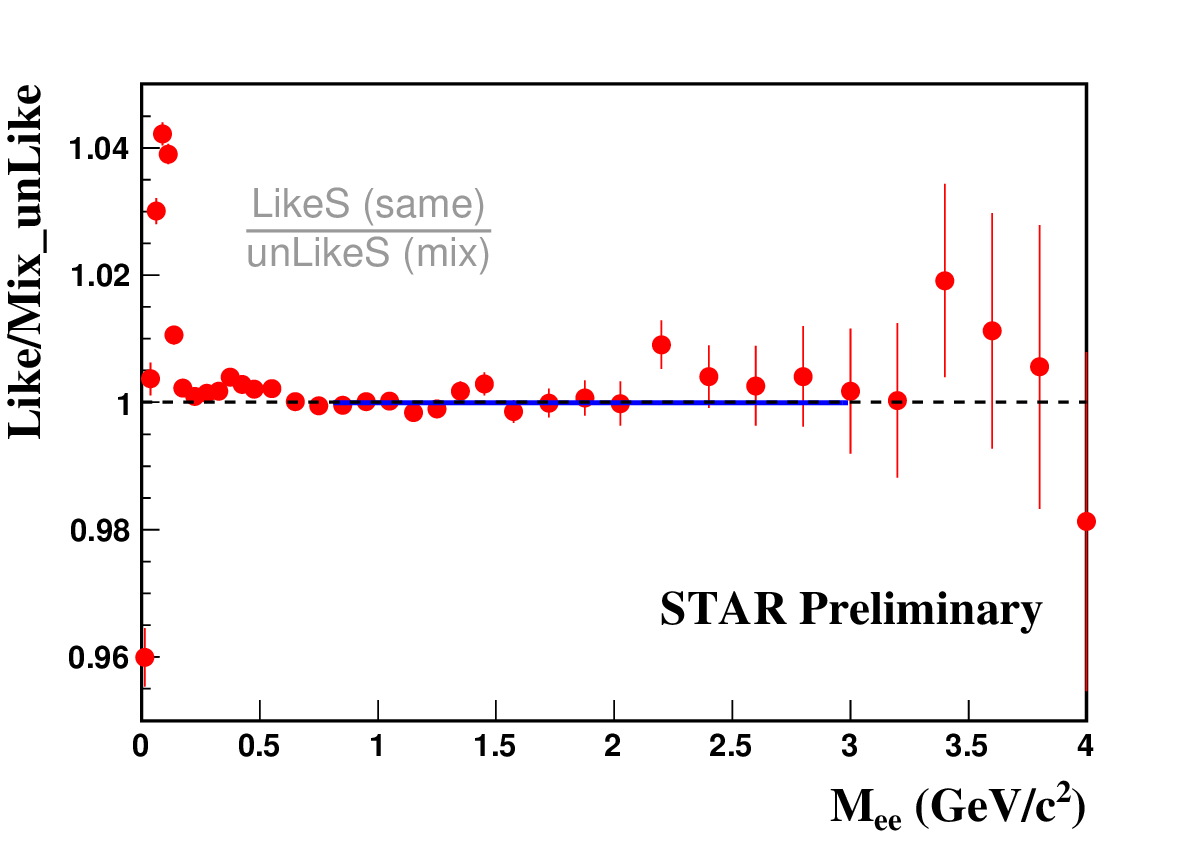}
    \caption{ Unlike-sign distribution in same and mixed event (left), and like-sign(same-event) to unlike-sign(mixed-event) ratio (right).} 
    \label{background} 
\end{figure}

The background reconstruction in this analysis is based on the mixed-event technique and the like-sign method. The invariant mass distribution from event mixing agrees with the same-event unlike-sign distribution in the IMR and HMR as shown in Fig.\ref{background}(left), and the right plot shows the ratio of the like-sign and mixed-event background. In the LMR, due to the correlated background, from {\it e.g.} cross pair and jet contribution\cite{PHEdie}, we subtract the like-sign background with an acceptance correction. The acceptance correction is needed because the acceptance of like sign pairs is slightly different to that of unlike sign pairs. In the IMR, we use the mixed-event background\cite{PHEdie} since it offers better statistics. The systematic error in the LMR is dominated by the acceptance uncertainty ($< 0.1\%$). In the IMR and HMR, it is dominated by the normalization factor uncertainty from different normalization methods and the normalization region ($< 0.1\%$). 
 
The dielecton continuum results in this paper are within STAR acceptance ($p_{T}^{e} > 0.2 GeV/c$ , $|\eta^{e}| < 1.0$ , $|y^{ee}| < 1.0$ ) and corrected for efficiency.   
\section{Results}
\begin{figure}[htbp!]
    \centering 
      \includegraphics[width=7.8cm]{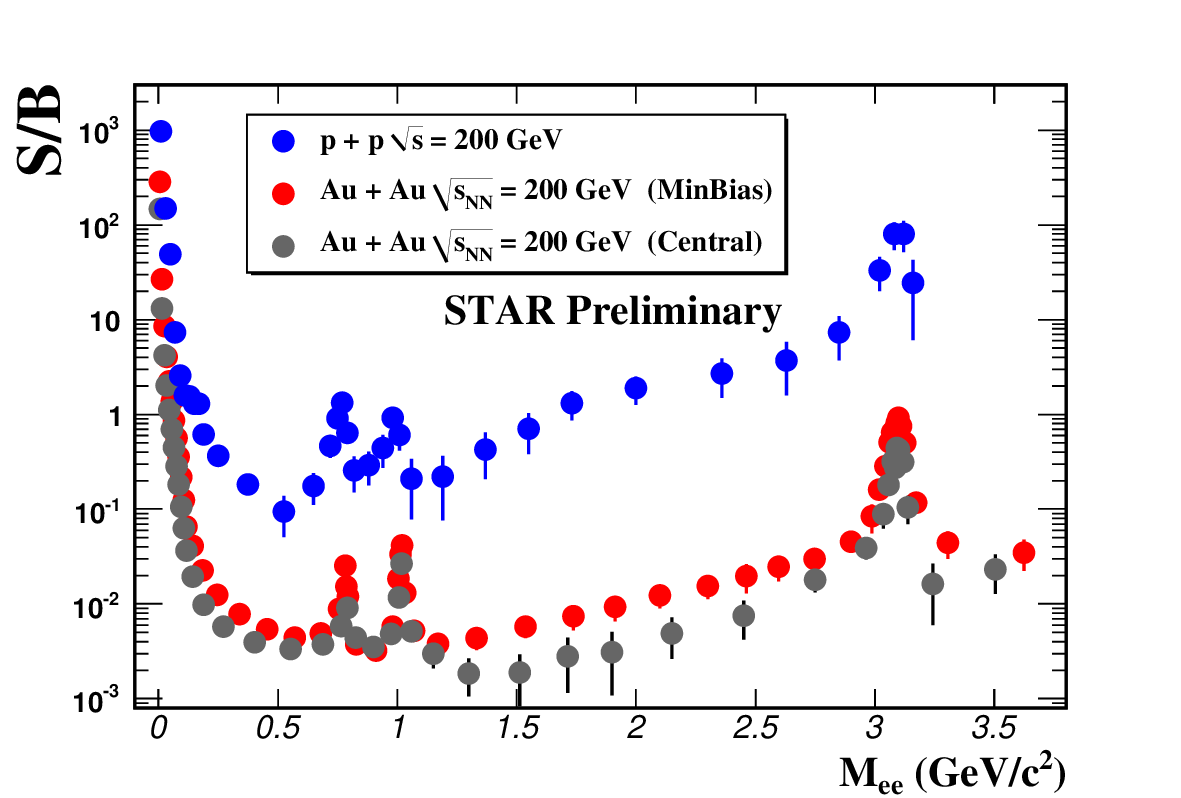} 
      \includegraphics[width=5.9cm]{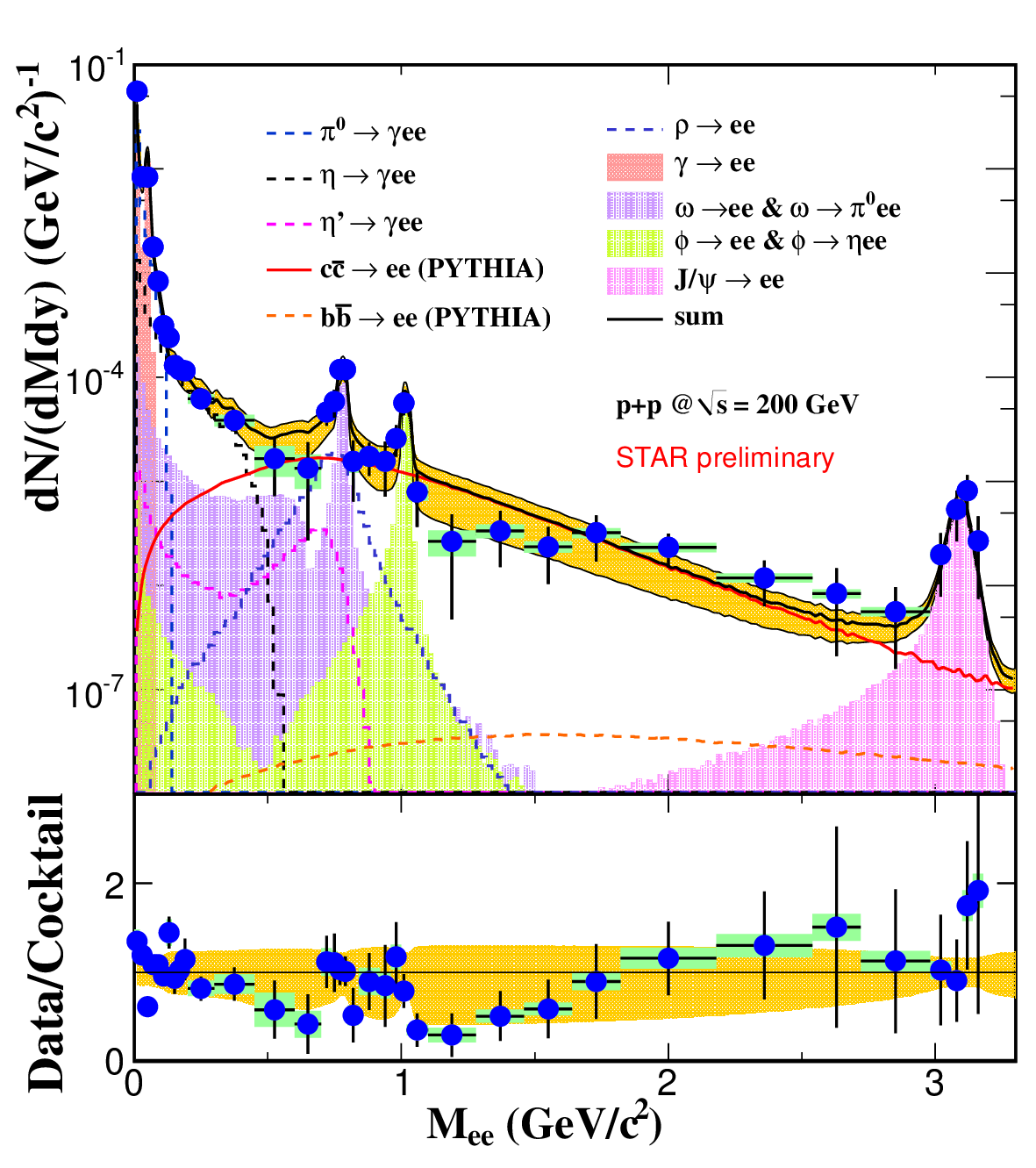} 
    \caption{ Signal/Background ratio (left), and invariant mass spectra from $\sqrt{s}$ = 200 GeV p + p collisions(right). The yellow band is systematic error on cocktail, and green boxes are systematic error on data.} 
    \label{pp} 
\end{figure}

\begin{figure}[htbp!]
    \centering 
      \includegraphics[width=5.9cm]{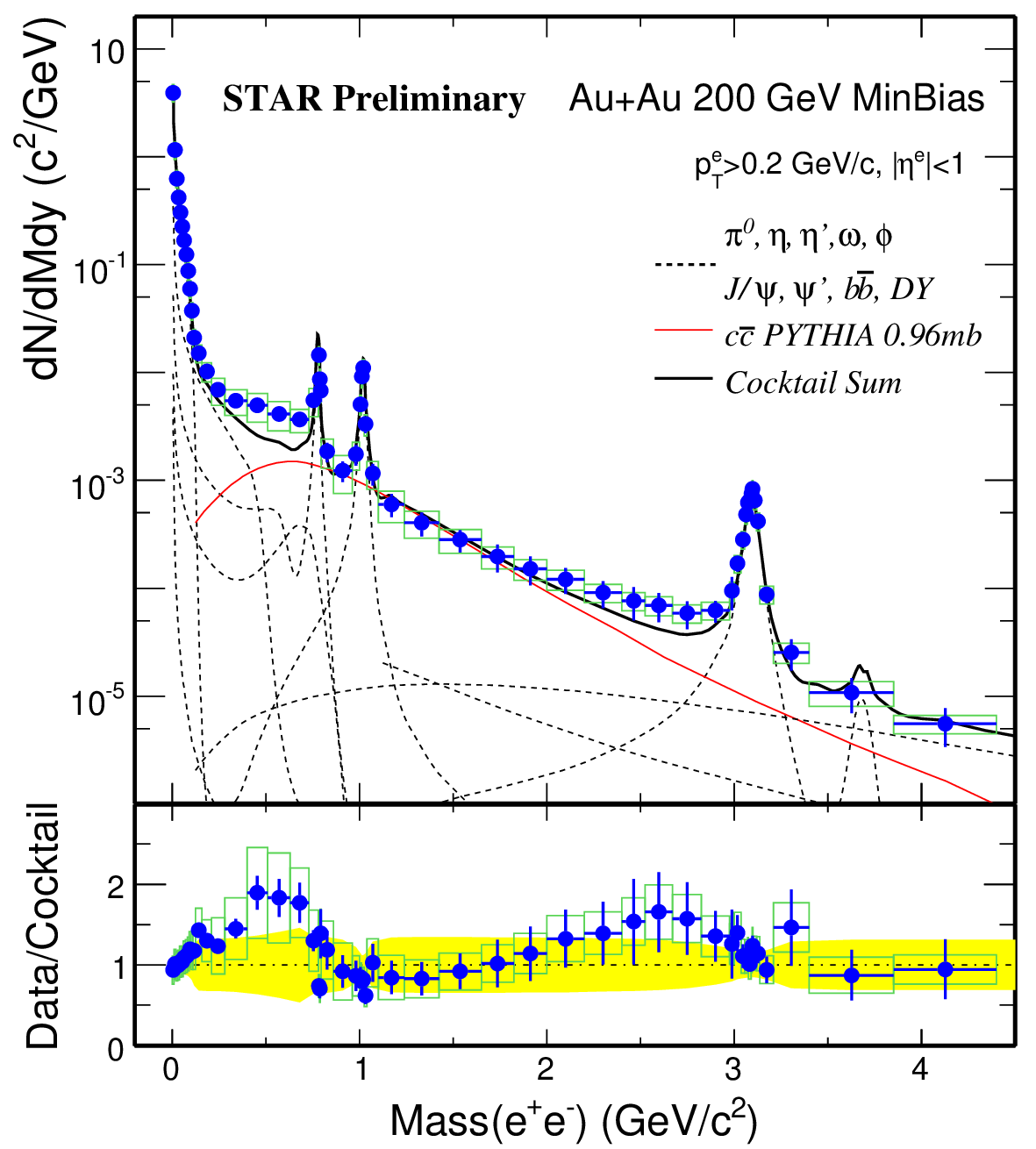} 
      \includegraphics[width= 5.9cm]{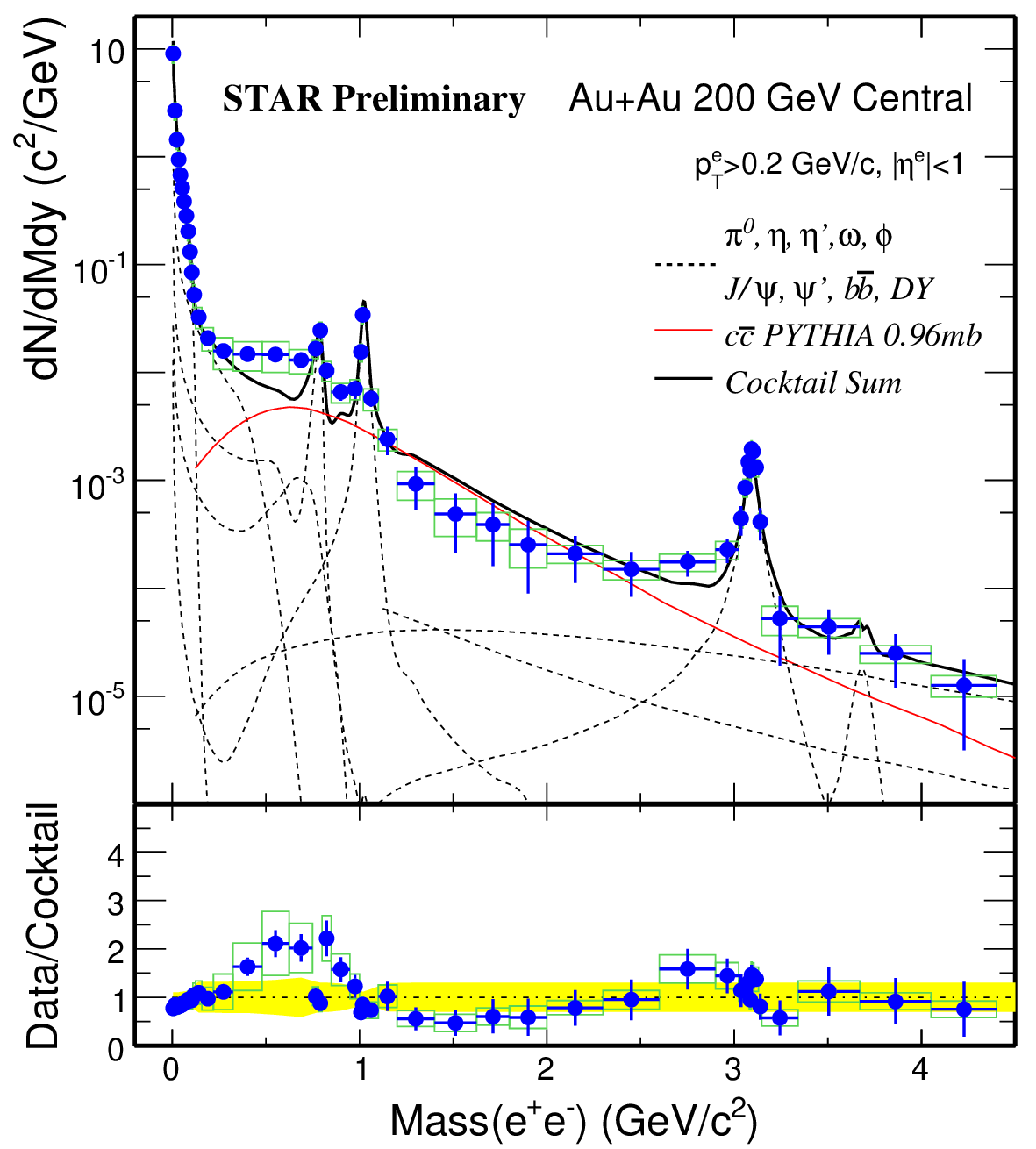} 
    \caption{ Invariant mass spectra from $\sqrt{s_{NN}}$ = 200 GeV Au + Au minimum-bias(left) and central collisions(right). The yellow band is systematic error on cocktail, and green box is systematic error on data.} 
    \label{AuAu} 
\end{figure}

\begin{figure}[htbp!]
    \centering 
      \includegraphics[width= 5.0cm]{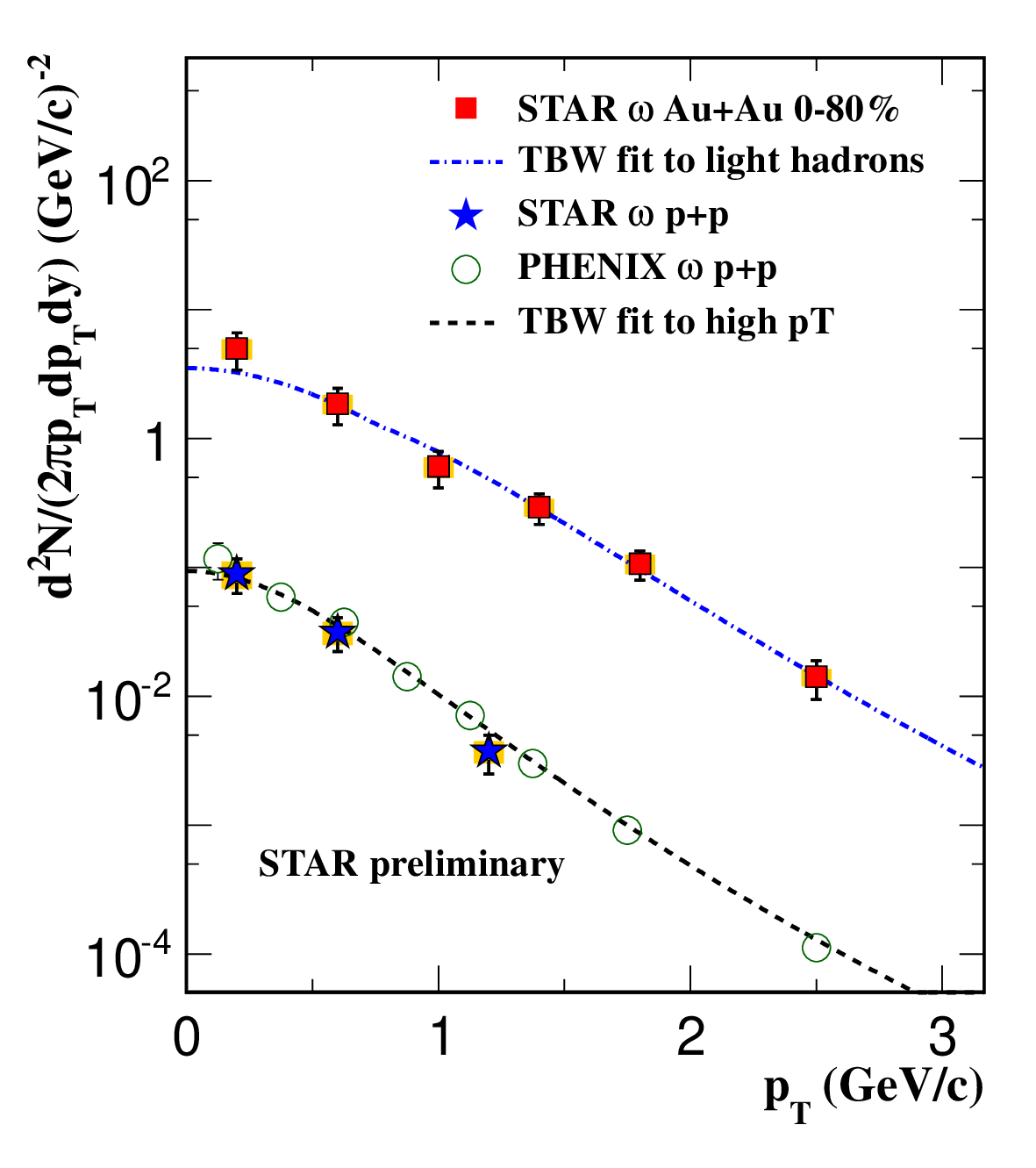} 
      \includegraphics[width=7.5cm]{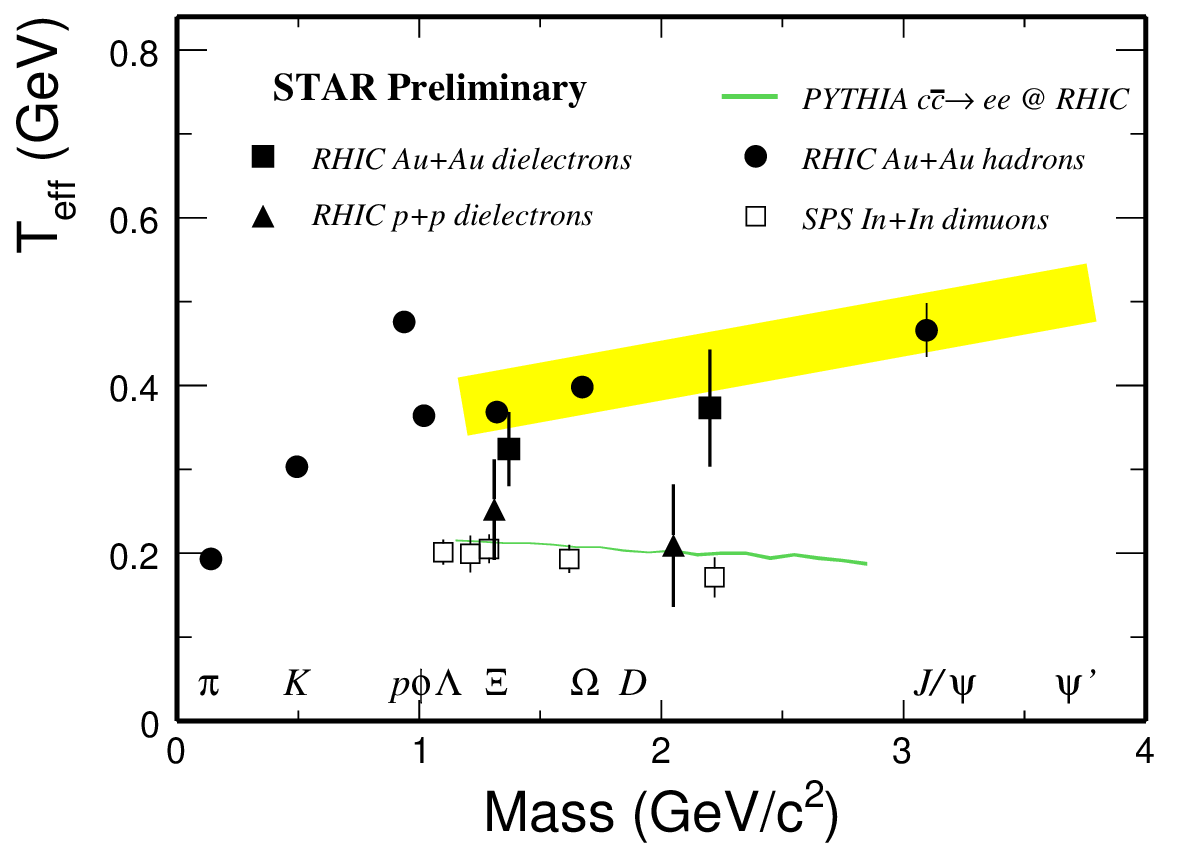} 
    \caption{$\omega$ $p_{T}$ spectra from $\sqrt{s_{NN}}$ = 200 GeV p + p and Au + Au collisions(left), and transverse mass slope parameters from RHIC and SPS measurements(right). } 
    \label{slope} 
\end{figure}

Figure \ref{pp} shows the Signal/Background ratio (left) and the invariant mass spectra from 200 GeV p + p collisions (right)\cite{Huang}.  Figure \ref{AuAu} shows the invariant mass spectra from 200 GeV Au + Au minimum-bias (left) and central collisions (right).
The $\omega \rightarrow e^{+}e^{-}$ spectra and the slope parameters at intermediate mass can be found in Fig.\ref{slope}\cite{Huang,NA60,STARh, PHEh}.

The results from p + p collisions are consistent with hadron decay cocktail simulations\cite{cocktail}, which provide a baseline for Au + Au collisions.

In the LMR, for central Au + Au collisions, we observer an enhancement with a factor of  $1.72 \pm 0.10^{stat} \pm 0.50^{sys} $ compare to cocktail (without $\rho$), for minimum-bias Au + Au collisions, the enhancement factor is $1.53 \pm 0.07^{stat} \pm 0.41^{sys} $ (without $\rho$). 

The measured $\omega \rightarrow e^{+}e^{-}$ spectra in p + p and Au + Au collisions are shown in Fig.\ref{slope}(left). The spectra in Au+Au collisions is consistent with the Tsallis Blave-Wave model prediction based on the fit to other hadrons \cite{tallis}. In p+p collisions, the spectra is consistent with the previous measurements \cite{PHEome} and Tsallis fit to high $p_{T}$ $\omega$ spectra \cite{tallis}.

In the IMR, compared to minimum-bias collisions, the yield from binary-scaled charm contributions from PYTHIA over-predicts the data in central collisions, which could indicate the modification of charm production in central Au + Au collisions. As show in Fig.\ref{slope}(right), the transverse mass slope parameters in p + p collisions are consistent with the PYTHIA charm, but they are different for the Au + Au results, as the slope parameters in Au + Au collisions are higher than those in p + p collisions. This may be due to thermal radiation and/or charm modification. We also compare our results with SPS results. Our inclusive dielectron slope parameters are higher than SPS dimuon results (charm/DY contribution subtracted) from NA60 Collaboration\cite{NA60}. 

In the future, the Heavy Flavor Tracker and Muon Telescope Detector upgrades will help us to separate the charm and thermal radiation contribution.

\section{Acknowledgments}
This work was supported in part by the NSFC of China under Grant
Nos. 11035009, the Shanghai Development
Foundation for Science and Technology under Contract No.
09JC1416800, and the Knowledge Innovation Project of the Chinese
Academy of Sciences under Grant No. KJCX2-EW-N01.


\section*{References}
\begin{thebibliography}{10}
\bibitem{TOF} Llope, et al, doi:10.1016/j.nima.2010.07.086;  TOF Collaboration, STAR TOF Proposal (2004)
\bibitem{TPC} Anderson, et al., Nucl. Instr. and Meth. A 499, 624 (2003)
\bibitem{PHEdie} PHENIX Collaboration, Phys. Rev. C 81, 034911 (2010)
\bibitem{Huang}B. Huang(for the STAR Collaboration), QM2011 poster
\bibitem{NA60} NA60 Collaboration, Phys. Rev. Lett 100, 022302 (2008)
\bibitem{STARh} STAR Collaboration, Nucl.Phys.A 757,102 (2005)
\bibitem{PHEh} PHENIX Collaboration, Phys. Rev. Lett 98, 232301 (2007)
\bibitem{cocktail}L. Ruan(for the STAR Collaboration), Nucl.Phys.A 855, 269 (2011)
\bibitem{tallis}Z. Tang et al., arXiv:1101.1912;  Z. Tang et al., Phys. Rev. C 79, 051901 (2009)
\bibitem{PHEome}PHENIX Collaboration, Phys. Rev. D 83, 052004 (2011).

\end{thebibliography}
\end{document}